\documentclass[twoside,a4paper,DIV=17]{scrartcl}

\usepackage{graphicx}
\usepackage[utf8]{inputenc}
\usepackage{url}
\usepackage{authblk}
\usepackage{hyperref}
\usepackage{cite}

\newcommand{\ep}{\Delta S_{\rm tot}}
\newcommand{\epb}{\tilde{\ep}}
\newcommand{\eprb}{\ep^{\rm B}}
\newcommand{\kb}{k_{\rm B}}
\newcommand{\ud}{{\rm d}}
\newcommand{\pf}{P}
\newcommand{\prb}{P_{\rm B}}
\newcommand{\pb}{\tilde P}
\newcommand{\ppf}{{\mathcal P}_{\rm F}}
\newcommand{\ppb}{{\mathcal P}_{\rm B}}
\newcommand{\bw}{_{\rm B}}
\newcommand{\traj}{{\bf m}}

\title{Fluctuation theorem between
non-equilibrium states in an $RC$ circuit}

\author[1,5]{Léo Granger}
\author[2,3]{Jumna Mehlis}
\author[3,5]{Édgar Roldán}
\author[4]{Sergio Ciliberto}
\author[3]{Holger Kantz}

\affil[1]{Departamento de F\'\i sica At\'omica, Molecular y Nuclear, 
Universidad Complutense de Madrid, 28040,  Madrid, Spain}
\affil[2]{
Martin-Luther-Universität Halle-Wittenberg, Von-Danckelmann-Platz
	3, D-06120, Halle, Germany}
\affil[3]{Max Planck Institut für die Physik komplexer Systeme,
Nöthnitzer Str.~38, D-01187, Dresden, Germany}
\affil[4]{Université de Lyon, Laboratoire de Physique, École Normale
Supérieure de Lyon, CNRS UMR5672, 46 allée d'Italie, 69364 Lyon Cedex
07, France}
\affil[5]{Grupo Interdisciplinar de Sistemas Complejos (GISC)}

\date{}

\begin{document}

\maketitle

\begin{abstract}
	Fluctuation theorems impose constraints on the probability of observing
	negative entropy production in small systems driven out of equilibrium.
	The range of validity of fluctuation theorems has been extensively
	tested for transitions  between equilibrium and non equilibrium
	stationary states, but not between general non equilibrium states. Here
	we report an experimental verification of the detailed fluctuation
	theorem for the total amount of entropy produced in the
	isothermal transition between two non-equilibrium states. The
	experimental setup is a parallel $RC$ circuit driven by an alternating
	current. We investigate the statistics of the heat released, of the
	variation of the entropy of the system, and of the entropy produced for
	processes of different durations. We show that the fluctuation theorem
	is satisfied with high accuracy for current drivings at different
	frequencies and different amplitudes.
\end{abstract}

\section{Introduction}

As already noted by Szilard in 1925 \cite{szilard_uber_1925},  entropy
reduction can occur in a single realization of a thermodynamic process at the
mesoscopic scale and the second law of thermodynamics is recovered when
averaging over many realizations of such a process. At scales where thermal
fluctuations are relevant, entropy-reducing trajectories can be observed
\cite{wang_experimental_2002,collin_verification_2005}.  The fluctuations of
the entropy production are governed by the so-called {\em fluctuation
theorems}, which relate the probability to observe a trajectory destroying a
certain amount of entropy to the probability to observe a trajectory producing
the same amount of entropy
\cite{evans_probability_1993,gallavotti_dynamical_1995,jarzynski_nonequilibrium_1997,crooks_nonequilibrium_1998,crooks_entropy_1999,hatano_steady-state_2001,seifert_entropy_2005,harris_fluctuation_2007,saha_entropy_2009,esposito_three_2010,spinney_fluctuation_2012}.
In particular, they ensure that on average, the entropy production is positive.
The fluctuation theorems are the building blocks of the emerging theory of
stochastic thermodynamics, which describes the equilibrium and non-equilibrium
thermodynamics of small systems, at the ensemble level as well as at the
trajectory level
\cite{seifert_entropy_2005,sekimoto_langevin_1998,esposito_second_2011,sekimoto_stochastic_2012,seifert_stochastic_2012,deffner_information_2012,granger_irreversibility_2013,roldan_irreversibility_2014,thingna_geometric_2014,bonanca_optimal_2014}.
In parallel to the theoretical development of stochastic thermodynamics, the
fluctuation theorems and the thermodynamics of small systems has been
intensively investigated experimentally in last decade
\cite{wang_experimental_2002,collin_verification_2005,liphardt_equilibrium_2002,trepagnier_experimental_2004,douarche_estimate_2005,tietz_measurement_2006,andrieux_entropy_2007,joubaud_fluctuations_2008,mestres_realization_2014,koski_distribution_2013,martinez_adiabatic_2015}.

The fluctuation theorem for the total entropy production relates the
probability  $\pf(\ep)$ to observe a trajectory producing an amount
$\ep$ of entropy in a given thermodynamic process to the
probability $\pb(-\ep)$ to observe a trajectory destroying the very
same amount of entropy in the time reversed or {\em backward} process,
where the driving of the system is reversed in time
\cite{crooks_entropy_1999,harris_fluctuation_2007,esposito_three_2010,spinney_fluctuation_2012}:
\begin{equation}
	\frac{\pf(\ep)}{\pb(-\ep)} =
	\exp\left(\frac{\ep}{\kb}\right).
	\label{e.ft}
\end{equation}
Equation~(\ref{e.ft}) has a wide range of validity: It is valid for
systems in contact with one or many heat baths, for transitions
between stationary or non-stationary states or for systems in
non-equilibrium stationary states.
This fluctuation theorem has been  experimentally tested
for the transition between equilibrium states
\cite{ritort_two-state_2002,collin_verification_2005}, where it
reduces to Crooks' relation \cite{crooks_nonequilibrium_1998},
for non-equilibrium steady-states
\cite{tietz_measurement_2006,joubaud_fluctuations_2008}, and in the
transition between non-equilibrium steady-states
\cite{trepagnier_experimental_2004}, where it can be refined to give the
Hatano-Sasa relation \cite{hatano_steady-state_2001}.
More recently, the fluctuation theorem (\ref{e.ft}) was observed in a more
general experiment involving a periodically driven system in contact with two
heat baths at different temperatures \cite{koski_distribution_2013}. All the
aforementioned experiments have this in common, that at the end of the backward
process, the system is in the same macroscopic state as at the beginning of
the forward process%
\footnote{ 
In fact, in the setup of \cite{koski_distribution_2013}, the macroscopic state
is time periodic and the succession of forward and backward processes
correspond to one period. The experiments described in
\cite{ritort_two-state_2002,collin_verification_2005,tietz_measurement_2006,trepagnier_experimental_2004}
involve transitions between (equilibrium or non-equilibrium) stationary
states.}.

In this paper, we report an experimental verification of the fluctuation
theorem (\ref{e.ft}) in a situation where this is not case: In our experiment,
the final state of the backward process is in general different from the state
the system was prepared in at the beginning of the forward process.  In such a
transition, the fluctuation theorem (\ref{e.ft}) presents a subtlety, as
realized by Spinney and Ford \cite{spinney_fluctuation_2012}. In fact, in
general, the distribution $\tilde P$ appearing in the denominator in the left
hand side of equation (\ref{e.ft}) \emph{is not the probability distribution of
the entropy produced in the backward process}.  In general, $\pb$ is the
distribution of a quantity which we will call \emph{conjugated entropy
production} in the following.  However, in situations were the final state of
the backward process is the same as the initial state of the forward process,
conjugated entropy production and entropy produced in the backward process are
equal.  Hence, in those cases, $\pb$ is the probability distribution of the
entropy produced in the backward process.

Our experimental system is a parallel $RC$ circuit driven by an alternative
current. We verify the fluctuation theorem (\ref{e.ft}) for processes of
arbitrary durations for different driving frequencies and intensities.
Furthermore, we show the difference between conjugated entropy production and
entropy produced in the backward process.

The paper is organized as follows.  We begin by sketching the derivation of the
fluctuation theorem (\ref{e.ft}) in section \ref{s.ft}. In the derivation, we
insist on the difference between the conjugated entropy production and entropy
produced in the backward process. We continue by describing the experimental
setup and protocol in section \ref{s.expSetup}. In particular, we show how we
sample forward and backward trajectories making the transition between two
non-equilibrium states.  Finally, we present our experimental results in
section \ref{s.res}. We study the statistics of the heat dissipated to the
environment, of the entropy produced and of the conjugated entropy production
for different driving times.  Furthermore, we show that the fluctuation theorem
(\ref{e.ft}) is satisfied for different driving speeds and amplitudes.  We
close the paper with a short discussion of our results in section \ref{s.ccl}.

\section{Detailed fluctuation theorem for the transition
between two non-equilibrium states}\label{s.ft}

\begin{figure}
\begin{centering}
	\includegraphics[scale=0.7]{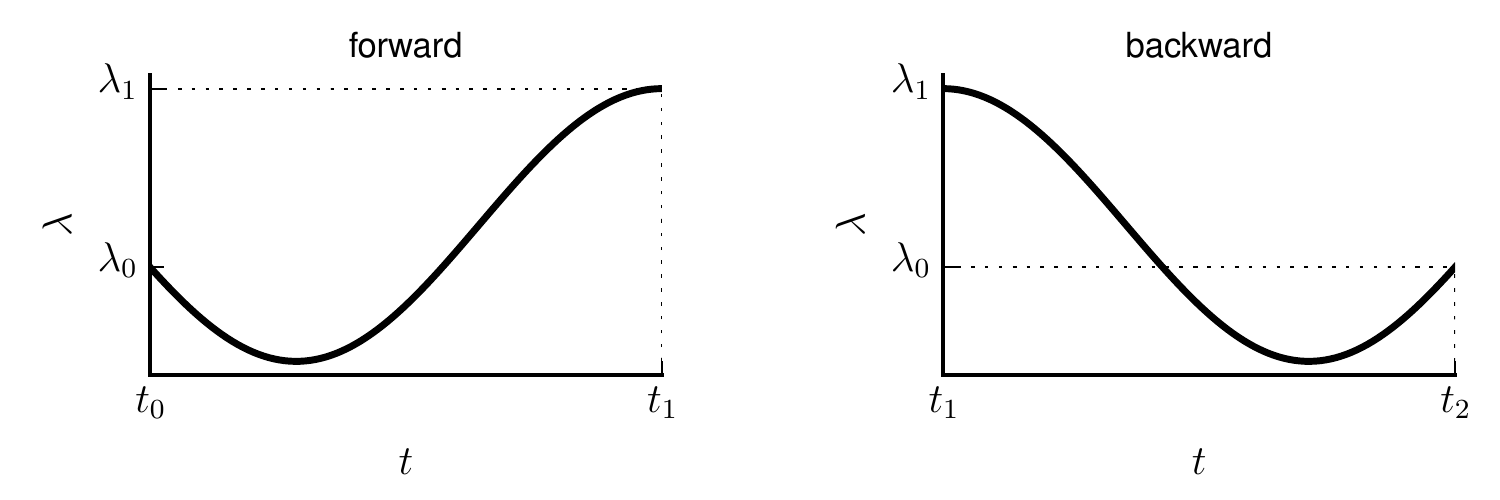}
	\caption{Forward (left panel) and time reverse (right panel)
	protocols.}
	\label{f.prots}
	\end{centering}
\end{figure}
We now sketch the derivation of equation (\ref{e.ft}) for the transition 
between two non-equilibrium states. Consider a small system in
contact with a heat bath at temperature $T$, that can be driven by
varying a control parameter $\lambda$.  Initially,
the value of the control parameter is set to $\lambda(t_0) =
\lambda_0$ and the system is prepared in a non-equilibrium macroscopic
state $\rho(q,t_0) = \rho_0(q)$. In other words, at time $t_0$, the
probability that the mesoscopic state of the system is $q$ is given by
the non-equilibrium distribution $\rho(q,t_0)$. Due to the presence of
thermal fluctuations, the mesoscopic state of the
system cannot be controlled, but only the {\em probability distribution} of mesoscopic
states. From time $t_0$ to $t_1$, the control parameter is 
changed from $\lambda(t_0) = \lambda_0$ to  $\lambda(t_1) =
\lambda_1$ according to a prescribed protocol $\lambda(t)$.  The final
macroscopic state of the system is $\rho(q,t_1) = \rho_1(q)$. In the
backward process, the initial state of the system is the final state
of the forward process, $\rho_1(q)$, and the control parameter takes
the same values as in the forward case but runs them backwards in
time. Hence, we assume that the time reverse process starts just after
the forward process ends, at time $t=t_1$.  The control parameter is
then varied according to $\lambda(t_1 + t) = \lambda(t_1 - t)$, as
sketched in figure \ref{f.prots}. The backward process ends at time $t_2
= t_1 + \tau$ where $\tau = t_1 - t_0$ is the duration of both the
forward and time reverse process. The final macroscopic state of the
system is $\rho(q,t_2) = \rho_2(q)$ which is in general different from
$\rho_0(q)$.

The amount of entropy produced by a trajectory $\traj= \{q_t : t_0
\le t \le t_1\}$ in the forward process is given by
\cite{kawai_dissipation:_2007,gomez-marin_footprint_2008,parrondo_entropy_2009,esposito_three_2010}:
\begin{equation}
	\ep[\traj] = \kb\log \frac{\ppf[\traj]}{\ppb[\bar\traj]},
	\label{e.ep}
\end{equation}
where $\ppf[\traj]$ is the probability to observe the trajectory
$\traj$ in the forward process and $\ppb[\bar{\traj}]$ is the
probability to observe the time reverse trajectory $\bar \traj$ of
$\traj$ in the backward process. The time reverse trajectory $\bar
\traj$ contains the same states as $\traj$, but runs backwards in
time: $\bar \traj(t_1 + t) = \traj(t_1 - t)$%
\footnote{We assume here that the system does not have degrees of
freedom that are odd under time reversal such as velocities. The
variables odd under time reversal should have their sign changed
in $\bar \traj$.}.
The definition of  entropy production  in (\ref{e.ep}) is consistent with the
usual thermodynamic definition
\cite{kawai_dissipation:_2007,gomez-marin_footprint_2008,parrondo_entropy_2009}:
\begin{equation}
	\ep[\traj] = \frac{Q[\traj]}{T} + \Delta S[\traj],
	\label{e.epthermo}
\end{equation}
where $Q[\traj]$ is the amount of heat dissipated to the environment
along the trajectory $\traj$ and $\Delta S[\traj] = -\kb\log \rho_1(q_1)
+ \kb \log \rho_0(q_0)$ is the variation of
the trajectory dependent entropy of the system along the trajectory $\traj$
\cite{seifert_entropy_2005}.
In fact, it can be shown that
\cite{seifert_entropy_2005,esposito_three_2010,andrieux_entropy_2007,kurchan_fluctuation_1998}:
\begin{equation}
	\frac{Q[\traj]}{T} = \kb \log
	\frac{\ppf[\traj|q_0]}{\ppb[\bar\traj|q_1]},
	\label{e.crooks}
\end{equation}
where $\ppf[\traj|q_0]$ is the probability to observe the trajectory
$\traj$ given the initial state $q_0$, and $\ppb[\bar \traj| q_1]$ is the
probability to observe the time reverse trajectory $\bar \traj$ in the
backward process given the initial state  $q_1$ of the time reverse
trajectory.
Equations (\ref{e.crooks}) and (\ref{e.epthermo}) together imply
(\ref{e.ep}).

Equation (\ref{e.ep}) can be rewritten as follows:
\begin{equation}
	{\ppf[\traj]} = {\ppb[\bar \traj]}\exp\left(
	\frac{\ep[\traj]}{\kb}
	\right).
	\label{e.ft2}
\end{equation}
For the time reverse process, let us {\em define}:
\begin{equation}
	\epb[\bar \traj] = \kb \log
	\frac{\ppb[\bar\traj]}{\ppf[\traj]} = -\ep[\traj].
	\label{e.epbw}
\end{equation}
With this definition,
integrating (\ref{e.ft2}) over all the trajectories that
produce the same amount of entropy $\ep$, we recover (\ref{e.ft}) with
\cite{esposito_three_2010,spinney_fluctuation_2012}:
\begin{eqnarray}
	\pf(\ep) &=&  \int \ppf[\traj]\, \delta\Big( \ep[\traj] - \ep
	\Big) \,\ud\traj,\\
	\pb(\ep) &=&  \int \ppb[\bar \traj]\, \delta\Big(
	\tilde{\ep}[\bar\traj] -
	\ep \Big) \ud\bar\traj.
	\label{e.pep}
\end{eqnarray}

However, $\tilde{\ep}[\bar\traj]$ defined in equation (\ref{e.epbw})
{\em is not the amount of entropy produced by the trajectory in the
time reverse process}. The latter is equal to
\begin{equation}
	\eprb[\bar \traj] = \frac{Q\bw[\bar \traj]}{T} + \Delta
	S\bw[\bar \traj],
	\label{e.eprb}
\end{equation}
where $Q\bw[\bar \traj]$ is the amount of heat dissipated to the
environment in the time reverse process along the trajectory $\bar
\traj$ and $\Delta S\bw[\bar \traj]$ is the variation of the entropy
of the system.
The heat released to the environment is odd under time reversal:
\begin{equation}
	Q\bw[\bar \traj] = \kb \log \frac{\ppb[\bar
	\traj|q_1]}{\ppf[\traj|q_0]} = -Q[\traj].
	\label{e.bwheat}
\end{equation}
However, this is not the case for the variation of the entropy of the system:
\begin{equation}
	\Delta S\bw[\bar \traj] = -\kb \log
	\frac{\rho_2(q_0)}{\rho_1(q_1)} \ne -\Delta S[\traj] = \kb \log
	\frac{\rho_0(q_0)}{\rho_1(q_1)}.
	\label{e.dsrb}
\end{equation}
Therefore, $\eprb[\bar \traj] \ne -\ep[\traj] = \epb[\bar \traj]$, and
the quantity defined in (\ref{e.epbw}) and entering the fluctuation
theorem  {\em is not the amount of entropy produced by the trajectory
$\bar \traj$ in the backward process}.

The expression of the entropy $\eprb[\bar \traj]$ produced in the backward
process in terms of probability of paths is:
\begin{equation}
	\eprb[\bar \traj] = \kb \log \frac{\ppb[\bar \traj |
	q_1]\rho_1(q_1)}{\ppf[\traj|q_0] \rho_2(q_0)}.
	\label{e.eprbpaths}
\end{equation}
In fact, while $\ppb[\bar \traj | q_1]\rho_1(q_1) = \ppb[\bar \traj]$
is the probability to observe the trajectory $\bar \traj$ in the
backward process, the probability to
observe the trajectory $\traj$ in the {\em time-reversal of the backward
process} is $\ppf[\traj|q_0] \rho_2(q_0)$.
In general, it is different from the probability to observe the
trajectory $\traj$ in the forward process: $\ppf[\traj|q_0]
\rho_2(q_0) \ne \ppf[\traj] = \ppf[\traj|q_0]\rho_0(q_0)$ because in
general the final macroscopic state of the backward process is not
equal to the initial macroscopic state of the forward process,
$\rho_2(q_0) \ne \rho_0(q_0)$.

We call $\epb[\bar \traj]$ the {\em conjugated entropy production}.  Using its
definition (\ref{e.epbw}) and the expression for the heat dissipated in the
backward process (\ref{e.bwheat}), we obtain that:
\begin{equation}
	\epb[\bar \traj] = \frac{Q\bw[\bar \traj]}{T} + \Delta \tilde
		S[\bar \traj],
	\label{e.epb}
\end{equation}
where
\begin{equation}
	\Delta \tilde{S}[\bar \traj] = -\kb \log
	\frac{\rho_0(q_0)}{\rho_1(q_1)} = -\Delta S[\traj].
	\label{e.dsb}
\end{equation}
Equations (\ref{e.epb}) and (\ref{e.dsb}) allow one to do a physical
interpretation of the conjugated entropy production.  The conjugated entropy
production is  equal to the variation of the entropy of the
environment in the backward process minus the variation of the entropy
of the  system in the forward process.

The conjugated entropy production is equal to the entropy produced in the backward process,
$\tilde{\ep}[\bar \traj] = \ep^{\rm B}[\bar \traj]$, if and
only if $\rho_0(q_0) = \rho_2(q_0)$,
i.e.~if the final macroscopic state of the backward
process is also the initial state of the forward process.
This is the case in the transition between equilibrium states and in
non-equilibrium stationary states. However, in the transition between
two arbitrary non-equilibrium, non stationary states, it has no reason
to be fulfilled.

The difference between $\epb[\bar \traj]$ and $\eprb[\bar \traj]$ is:
\begin{equation}
	\epb[\bar \traj] - \eprb[\bar \traj] = \Delta \tilde S[\bar \traj]
	- \Delta {S}\bw[\bar \traj] = \kb \log
	\frac{\rho_2(q_0)}{\rho_0(q_0)}.
	\label{e.diff}
\end{equation}
When averaging over many trajectories of the backward process,
this difference is positive:
\begin{equation}
	\langle \epb \rangle - \langle \eprb \rangle = \kb \int
	\rho_2(q_0) \log \frac{\rho_2(q_0)}{\rho_0(q_0)}\ud q_0 \ge 0.
	\label{e.meandiff}
\end{equation}
The right hand side of~(\ref{e.meandiff})  is the {\em relative
entropy} or {\em Kullback-Leibler divergence} between the
distributions $\rho_2$ and $\rho_0$. This quantity is non negative and
it is zero if and only the two distributions $\rho_2$ and $\rho_0$ are
undistinguishable \cite{cover_elements_1991}.

\section{Experimental setup}\label{s.expSetup}

\subsection{The system}

\begin{figure}
\begin{centering}
	\includegraphics{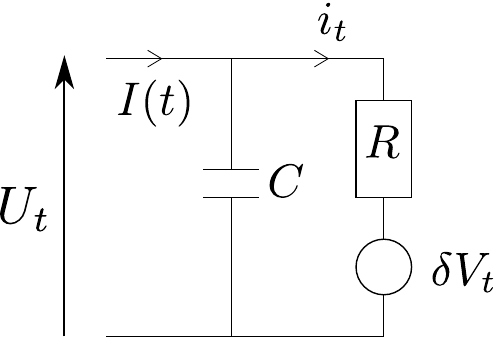}
	\caption{Sketch of the experimental setup. A resistor of $R = 1 \,{\rm
	M\Omega}$ is connected in parallel with a capacitor of $C = 1\,{\rm
	nF}$. The voltage source $\delta V_t$ in series with the resistor
	represents Johnson-Nyquist noise.
	}
	\label{f.circuit}
	\end{centering}
\end{figure}
The experimental setup is sketched in figure \ref{f.circuit}.  A resistor of
resistance $R = 1\,{\rm M\Omega}$ is connected in parallel with a condenser of
capacity $C = 1 \,{\rm nF}$. The input current $I(t)$ is oscillating at a
frequency of $f_{\rm d}$, $I(t) = I_{\rm max} \sin(\omega t)$ with  $\omega =
2\pi f_{\rm d}$.  The time constant of the circuit is $\tau_{\rm c} = RC
\approx 1\,{\rm ms}$.

The voltage across the resistor fluctuates due to Johnson-Nyquist noise,
which is modelled in figure
\ref{f.circuit} by putting a voltage source in series with the
resistor. At any time, this source produces a random voltage $\delta
V_t$ satisfying \cite{van_zon_power_2004,garnier_nonequilibrium_2005}:
\begin{eqnarray}
	\langle \delta V_t\rangle &=& 0\\
	\langle \delta V_t \delta V_{t'} \rangle &=&
	2{\kb T}{R}\delta(t - t').
	\label{e.johnson}
\end{eqnarray}
We denote $q_t$ the charge that has flown through the resistor at time
$t$ and $i_t = \ud q_t/\ud t$ is the current that flows through
it. Moreover, let $q^*_t = \int_{-\infty}^t I(s) \ud s$ the total
charge that has flow through the circuit at time $t$.
The charge of the capacitor is thus $q_t^* - q_t$ and the voltage
across the circuit is equal to:
\begin{equation}
	U_t = \frac{q_t^* - q_t}{C}.
	\label{e.cond}
\end{equation}
Ohm's law for the resistor implies that:
\begin{equation}
	U_t = R i_t + \delta V_t.
	\label{e.ohm}
\end{equation}
Hence, the charge $q_t$ obeys the following Langevin equation:
\begin{equation}
	R \frac{\ud q_t}{\ud t} = - \frac{1}{C}\left( q_t - q^*_t
	\right) + \delta V_t.
	\label{e.langevin}
\end{equation}
This equation is identical to the equation of motion of an overdamped
Brownian particle whose position is $q_t$, its friction coefficient is
$R$ and  is trapped with a harmonic trap of stiffness $1/C$ centered
at $q_t^*$. Our control parameter is $q^*_t$. It oscillates sinusoidally at
frequency $f_{\rm d}$ and its amplitude is related to the amplitude of the
input current through $q_{\rm max}^* = I_{\rm max} / \omega$.
\begin{figure}
\begin{centering}
	\includegraphics[scale=1]{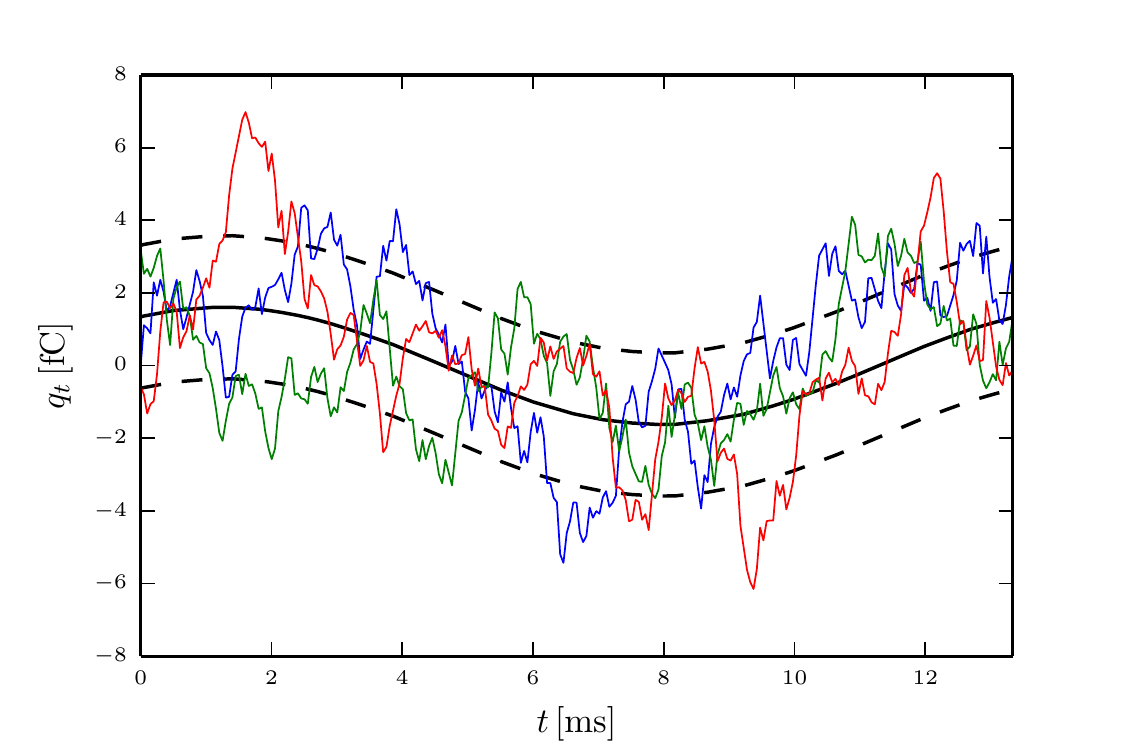}
	\caption{Three example of stochastic trajectories during one
		period of the driving signal (red, green and blue
		strongly fluctuating lines). The black solid line
		represents the ensemble average $\langle q_t \rangle$
		and the black dashed lines represent the ensemble
		average plus and minus one standard deviation
		$\sigma_t = \sqrt{\left\langle( q_t - \langle q_t
		\rangle )^2 \right\rangle}$. In this example, the driving
		frequency is $f_{\rm d} = 75 \,\textrm{Hz}$ and the amplitude
		of the driving is $q_{\rm max}^* = 2.3\,\textrm{fC}$. The
		average and standard
		deviation are estimated from 38164 trajectories.
		}
	\label{f.traj}
	\end{centering}
\end{figure}
In figure \ref{f.traj}, we plot three examples of realizations of the
stochastic trajectory $q_t$, together with the ensemble average.

\subsection{Protocol}

\begin{figure}
	\includegraphics{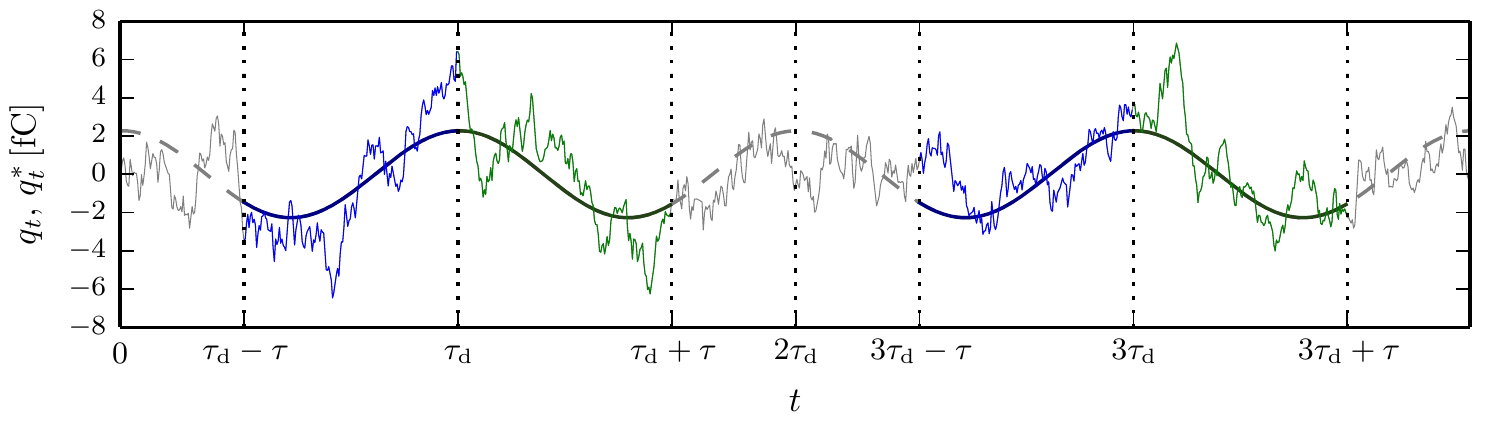}
	\caption{Construction of the ensemble of non-equilibrium
	trajectories for a process duration of $\tau = 8.4 \,{\rm ms}$
	from a long stationary trajectory driven at $f_{\rm d} =
	75\,\textrm{Hz}$. Since $\tau < 1/f_{\rm d} = 13.3\,\textrm{ms}$, we
	set $\tau_{\rm d} = 1/f_{\rm d}$.
	The smoothly
	oscillating thick curve
	represents the control parameter $q_t^*$. The strongly
	fluctuating thin curve represents $q_t$. The blue portions
	correspond to members of the forward ensemble and the green
	portions to members of the backward ensemble.  The forward
	trajectories start at time $(2n+1) \tau_{\rm d} - \tau$ and end at
	time $(2n+1) \tau_{\rm d} $, and the backward
	trajectories start at time $(2n+1)\tau_{\rm d}$ and end at
	time $(2n+1)\tau_{\rm d} -\tau$, where $n$ is an integer
	(here, we plot $n = 0$ and
	$n = 1$).  Due to the periodicity of the driving protocol,
	each member of the forward ensemble is subject to the same
	driving signal and due to the symmetry of the driving signal,
	the protocol in each backward process is the time reverse of
	the protocol in the forward process.
	}
	\label{f.expprot}
\end{figure}
After a transient that we do not analyze here, the system relaxes
towards a time periodic stationary state. In other words, the probability
distribution $\rho(q,t)$ of the charge $q$ at time $t$ is time
periodic of period $1/f_{\rm d}$, the period
of the driving signal.  We use this periodicity of $\rho(q,t)$ to
construct an ensemble of non-equilibrium trajectories.

Here is how we construct the ensembles of forward and backward trajectories of
duration $\tau$ from a long quasistationary trajectory $\{q_t\}$.  We chose the
origin of time such that $q_t^* = q^*_{\rm max} \cos(\omega t)$, where $\omega
= 2\pi f_{\rm d}$. Let $\tau_{\rm d} \ge \tau$ be an integer multiple of the driving
period $1/f_{\rm d}$. The $n^{\rm th}$ member of the forward ensemble is the
portion of $\{q_t\}$ where $  (2n + 1) \tau_{\rm d} -\tau \le t \le
(2n+1)\tau_{\rm d}$, $n$ being an integer. The corresponding member of the
backward ensemble is the portion of $\{q_t\}$ where $  (2n+1)\tau_{\rm d}\le t
\le(2n + 1) \tau_{\rm d} +\tau $.

On figure \ref{f.expprot}, we sketch how the first two members of the forward
and backward ensembles are obtained for $\tau_{\rm d} = 1/f_{\rm d}$.
The blue portions correspond to members of
the forward ensemble and the green portions to members of the backward
ensemble. The smoothly oscillating curve represents the driving signal $q^*_t$
and the strongly fluctuating curve is $q_t$.

Due to the
periodicity of the driving protocol, all the members of the forward
ensemble are subjected to the same driving signal.
Moreover, since $\rho(q,t)$ is time periodic of period $\tau_{\rm d}$, all the
initial mesoscopic states of the forward trajectories are drawn from the same
initial distribution $\rho_0(q)$ and hence all the forward trajectories are
drawn from the same path distribution $\ppf$.
The same reasoning applies the
backward trajectories: their initial mesoscopic state is drawn from the same
distribution $\rho_1(q)$ and they are all submitted to the same driving. Hence
they are all drawn from the same path distribution $\ppb$.
Finally, the origin of time was chosen such that $q_t^*$ is symmetric around
$n \tau_{\rm d}$, $q^*_{n\tau_{\rm d} - t} = q^*_{n\tau_{\rm d} + t} $.
Hence the members of the backward ensemble are subjected to a driving
that is the time reversal of the driving under which the members of the forward
ensemble are submitted.

\subsection{Measurement of stochastic entropy production}

The amount of entropy produced along a stochastic trajectory $\traj$
is calculated using (\ref{e.epthermo}): $\ep[\traj] = Q[\traj] / T + \Delta
S[\traj]$. Following Sekimoto \cite{sekimoto_stochastic_2012}, the heat  released
to the environment in the time interval $[t,t+dt]$ is given by
\begin{equation} 
	\ud Q_t = \left(  R\frac{\ud q_t}{\ud t}  + \delta V_t \right)\circ
	\ud q_t = -\frac{1}{C}\left(q_t - q_t^{\star} \right) \circ \ud q_t,
	\label{e.heat}
\end{equation}
where $\circ$ denotes Stratonovich product and the second equality is
consequence of (\ref{e.langevin}).  Note that the amount of heat
released per unit time is the sum of two contributions, $\dot Q_t =
\dot Q_t^{\rm Joule} + \dot Q_t^{\rm thermal}$.  The first contribution is
due to the Joule heating inside the resistor: $\dot Q_t^{\rm Joule} =
R i_t^2$, and the second to the power injected by thermal
fluctuations: $\dot Q_t^{\rm thermal} = \delta V_t i_t$.
The heat dissipated between
times $t_0$ and $t_1$ along a stochastic trajectory $\traj$ equals to
\begin{equation}
	Q[\traj] = \int_{t_0}^{t_1} \ud Q_t  =
	-\int_{t_0}^{t_1} \frac{q_t - q_t^*}{C} \circ \ud q_t,
	\label{e.heattraj}
\end{equation}
which is measured from the stochastic trajectories.

The trajectory dependent entropy is given by $S(q_t,t) = -\kb\log
\rho(q_t,t)$, where $\rho(q_t,t)$ is the probability distribution of
the charge $q_t$ at time $t$.  The distribution $\rho(q_t, t)$ is
estimated from the ensemble of trajectories.  The system's entropy
change along a trajectory $\traj$ that starts at $q_0$ at time $t_0$
and ends at $q_1$ at time $t_1$ is obtained as
\begin{equation} \Delta S[\traj] = S(q_1,t_1) -
	S(q_0, t_0) = -\kb \log \rho(q_1, t_1) + \kb \log
	\rho(q_0,t_0).  \label{e.trajent}
\end{equation}

\section{Experimental results}\label{s.res}

\begin{figure}
	\begin{centering}
	\includegraphics[scale = 1.00]{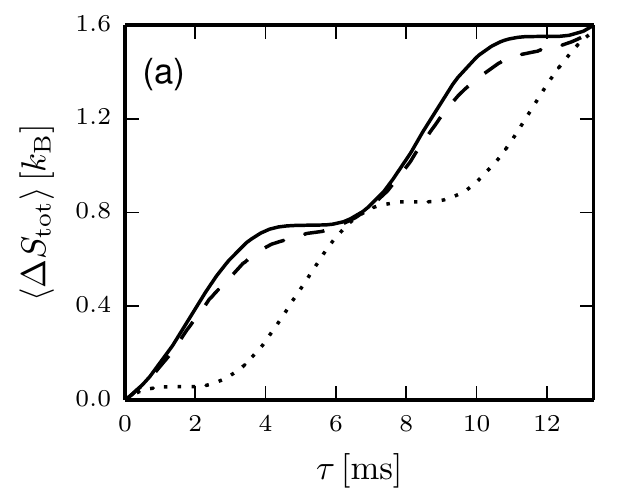}
	\hspace{0.16cm}
	\includegraphics[scale = 1.00]{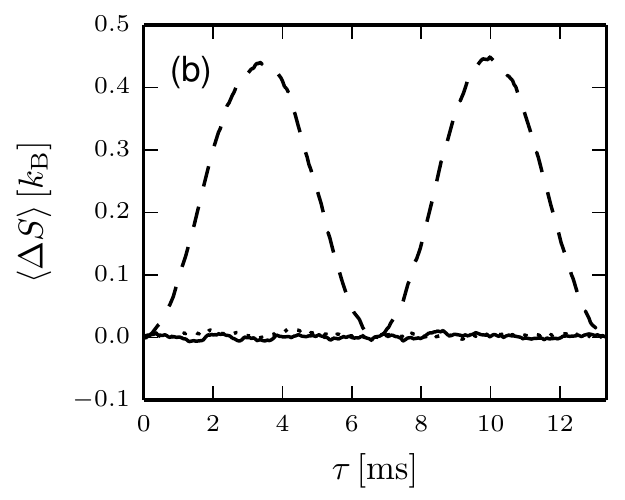}
	\caption{(a) Average entropy production as a function of the
		process duration $\tau$. The solid line represents the
		entropy production in the forward process, the dotted
		line the entropy production in the
		backward process and the dashed line the conjugated entropy production
		$\tilde{\ep}[\bar \traj]$. (b) Average variation of the entropy of the
		system in the forward (solid line) and in the time
		reverse process (dotted line). They are both equal to
		zero for all $\tau$. The dashed line represents the
		average of $\Delta \tilde{S}[\bar \traj]$ over the
		time reverse process.
		}
	\label{f.mep}
	\end{centering}
\end{figure}

\begin{figure}
	\includegraphics[scale=1.00]{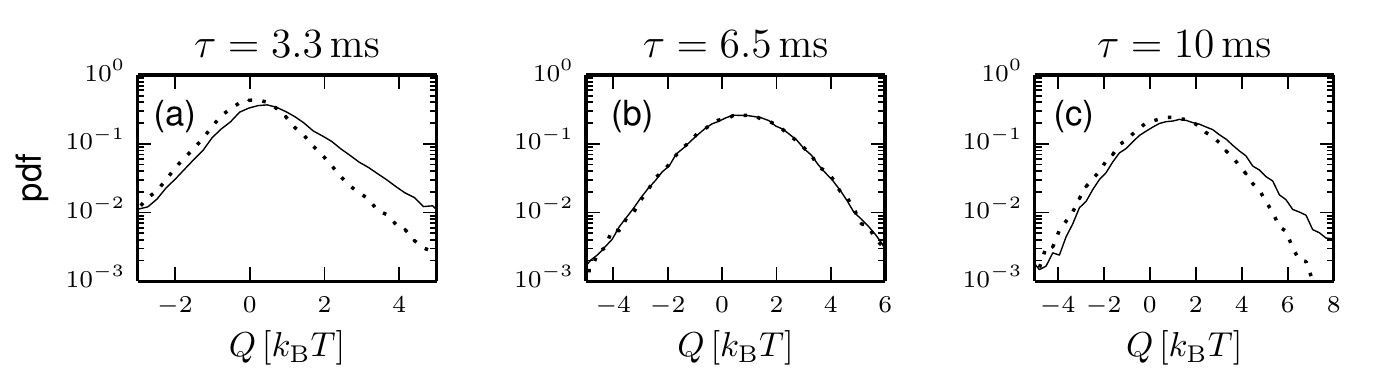}

	\includegraphics[scale=1.00]{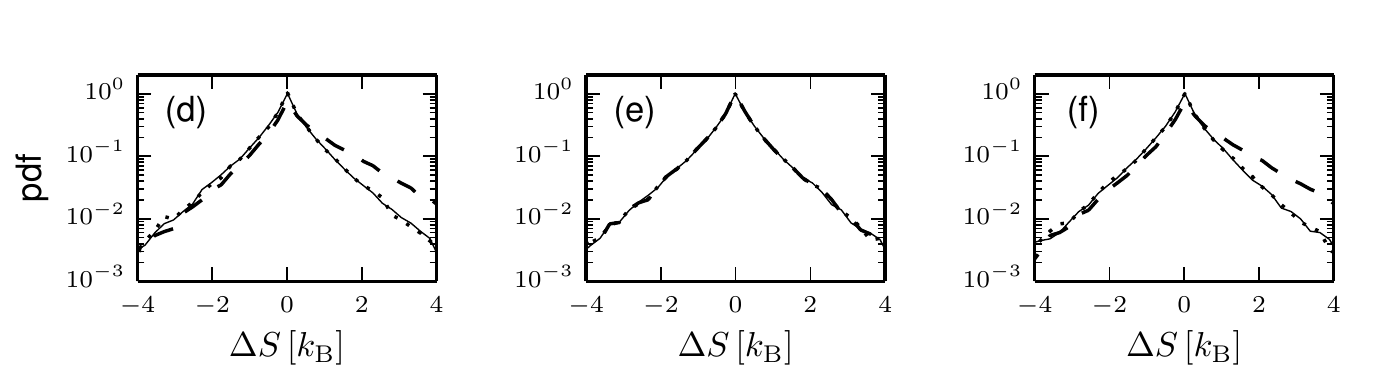}

	\includegraphics[scale=1.00]{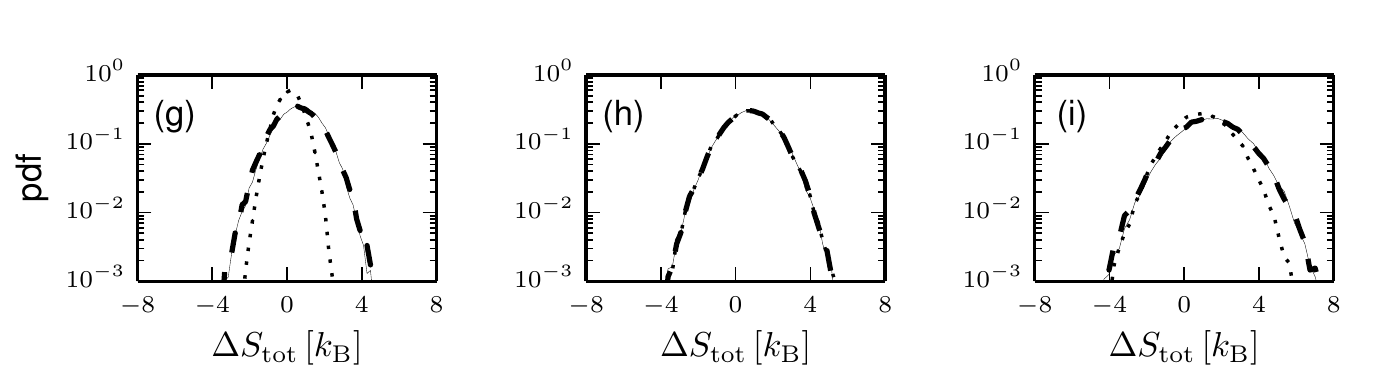}
	\caption{Distributions of the heat (top), system entropy variation (middle) and
		entropy production (bottom) for the three different durations
		$\tau \approx 3.3$, $6.5$ and $10\,{\rm ms}$
		corresponding respectively to one quarter, one half
		and three quarters of the driving period $\tau_{\rm
		d}$. Each column corresponds to one value of $\tau$.
		Upper row (panels a, b, and c): Distributions of the
		heat dissipated in the forward (solid lines) and time
		reverse process (dashed lines). Middle row (panels d,
		e, and f): Distributions of the variation of the
		entropy of the system in the forward process (solid
		lines), in the time reverse process (dotted lines) and
		of $\Delta \tilde{S}[\bar \traj]$ given by
		(\ref{e.dsb}) in the time reverse process (dashed
		lines). Lower row (panels g, h, and i): Distribution
		$\pf(\ep)$ of the entropy produced in the forward
		process (solid lines), distribution $\prb(\ep)$ of the
		entropy produced in the backward process (dotted
		lines), and distribution $\pb(\ep)$ of the conjugated
		entropy production (dashed lines).
	}
	\label{f.hists}
\end{figure}

Figures \ref{f.mep} and \ref{f.hists} summarize the thermodynamics of the
process as a function of the process duration $\tau$ for a driving frequency of
$f_{\rm d}  = 75\,\textrm{Hz}$, and hence a driving period of $1/f_{\rm d}
\approx 13.3\,\textrm{ms}$, and a driving amplitude of $q^*_{\rm max} =
2.3\,\textrm{fC}$. We consider durations up to one period of the driving signal
and hence we set $\tau_{\rm d} = 1/f_{\rm d}$.
The signal $q_t$ was sampled at $20\,\textrm{kHz}$ and the
ensemble consists of 38164 trajectories.

Figure \ref{f.mep}a shows the ensemble averages of the amount of
entropy produced in the forward process, $\langle \ep \rangle$ (solid
line), the amount of entropy produced in the backward process
$\langle \eprb \rangle$ (dotted line) and of the conjugated entropy
production $\langle
\epb \rangle$ (dashed line) as a function of the process duration
$\tau$.  These three quantities increase with $\tau$, in a manner that
is roughly linear with a periodic modulation.  We check the inequality
(\ref{e.meandiff}), $\langle \epb \rangle \ge \langle \eprb \rangle$.
Moreover, the average amount of entropy produced in the forward
process is approximately equal to the average conjugated entropy production,
$\langle \ep \rangle \approx \langle \epb \rangle$.

We also investigate the average value of the system's entropy change (figure
\ref{f.mep}b) in the forward process, $\langle \Delta S \rangle$
(solid line), in the time reverse $\langle \Delta S\bw\rangle$ (dotted
line) and the quantity $\langle \Delta \tilde{S}\rangle$ (dashed line)
as a function of the process duration $\tau$.  The 
system's entropy change vanishes on average both in forward and backward
processes.  Hence, the average entropy production is equal to the
average dissipated heat in both cases, $\langle \ep \rangle = \langle
Q \rangle/T$ and $\langle \eprb \rangle = \langle Q\bw\rangle/T$.

In this situation (\ref{e.diff}) and (\ref{e.meandiff}) imply
\begin{equation}
	\langle \epb \rangle - \langle \eprb \rangle = \langle \Delta
	\tilde{S}\rangle = \kb \int \rho_2(q) \log
	\frac{\rho_2(q)}{\rho_0(q)}\,\ud q \ge 0.
	\label{e.meandiffexp}
\end{equation}
On figure \ref{f.mep}b we can see that $\langle \Delta\tilde{S}\rangle$ is non
negative for all $\tau$, in accordance with (\ref{e.meandiffexp}).  The
quantity $\langle \Delta \tilde{S}\rangle$ is zero for $\tau = 0$, $\tau =
\tau_{\rm d}/2 \approx 6.5\,{\rm ms}$ and  $\tau = \tau_{\rm d}$, implying that
for these durations, $\rho_2 \equiv \rho_0$.  In fact, for $\tau = 0$, we have
$t_0 = t_1 = t_2$ and there is no process and hence $\rho_0 \equiv \rho_1
\equiv \rho_2$.  For $\tau = \tau_{\rm d} / 2$, we have $t_2 - t_0 = \tau_{\rm
d}$, and hence $\rho_2(q) = \rho(q,t_0+\tau_{\rm d}) = \rho(q,t_0) = \rho_0(q)$
because $\rho(q,t)$ is time periodic with period $\tau_{\rm d}$. The same
reasoning applies for $\tau = \tau_{\rm d}$. In that case, we have $t_2 - t_0 =
2\tau_{\rm d}$, and hence $\rho_2(q) = \rho(q,t_0+2\tau_{\rm d}) = \rho(q,t_0)
= \rho_0(q)$.  The quantity $\langle \Delta \tilde{S}\rangle$ is also zero for
$\tau = \tau_{\rm d}/2 \approx 6.5\,{\rm ms}$ which is one half of the driving
period.  Finally, $\langle \Delta \tilde{S}\rangle$ is maximum for
$\tau \approx 3.3\,{\rm ms}$ and $\tau \approx 10\,{\rm ms}$ which correspond
to one fourth and three fourth of the driving period.

On figure \ref{f.hists}, we show the distributions of the
thermodynamic quantities heat, entropy variation and entropy
production in the forward and backward process for the three durations
$\tau \approx 3.3$, $6.5$ and $10\,{\rm ms}$, corresponding to one
quarter, one half and three quarters of the period of the driving
signal.

The first row of figure \ref{f.hists} (panels a, b, and c) shows the
distributions of the heat released to the environment in  the forward
(solid lines) and in the time reverse process (dashed lines). These
are identical for $\tau \approx 6.5\,{\rm ms}$ otherwise they are
different.

In the middle row (panels d, e and f), we show
the distributions of the system's entropy change in the forward
process (solid lines) in the backward process (dotted lines) and of
the quantity $\Delta \tilde{S}$ (dashed lines). The distributions of
the system's entropy change in forward and backward processes are
identical. They are symmetric with respect to $0$ and non-Gaussian for
all values of $\tau$. Similar results were also found in
\cite{joubaud_fluctuations_2008,martinez_adiabatic_2015} The
distribution of $\Delta \tilde{S}[\bar \traj]$ differs from the two
others except for $\tau \approx 6.5\,{\rm ms}$.

The lower row (panels g, h and i) of figure \ref{f.hists} shows the
distribution $\pf(\ep)$ of the entropy produced in the forward process
(solid lines), the distribution $\prb(\ep)$ of the entropy produced in
the backward process (dotted lines) and the distribution $\pb(\ep)$ of
the conjugated entropy production (dashed lines). The three distributions are
Gaussian, as in \cite{joubaud_fluctuations_2008}. The distributions
of the entropy produced in the forward process and of the conjugated entropy
production
are equal, $\pf(\ep) = \pb(\ep)$.
The distribution $\prb(\ep)$ of the entropy produced in the backward
process is equal to the two others for $\tau = 6.5\,{\rm ms}$,
otherwise it has a different shape. In accordance with equation
(\ref{e.meandiff}) and with figure \ref{f.mep}a, its mean is smaller
than the mean of the two others. Moreover, its variance is also
smaller. Note that a necessary condition for the fluctuation theorem
(\ref{e.ft}) to hold when the distributions $\pf(\ep)$ and $\pb(\ep)$ are
Gaussian is that they are equal
\cite{douarche_estimate_2005,granger_crooks_2010}.

\begin{figure}
	\includegraphics{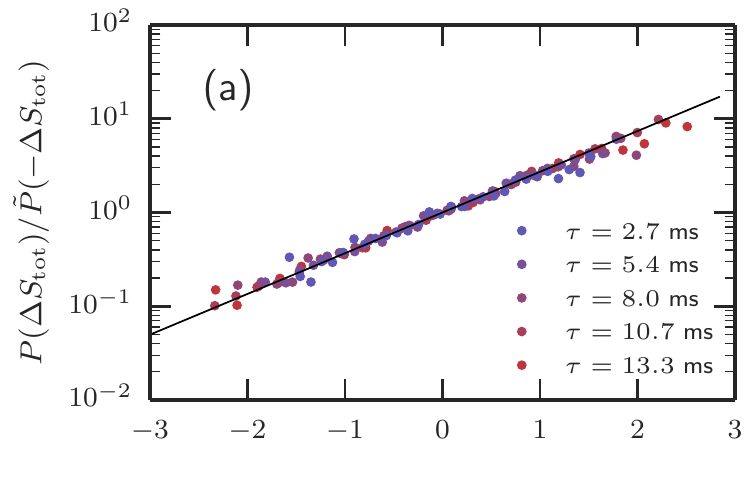}
	\includegraphics{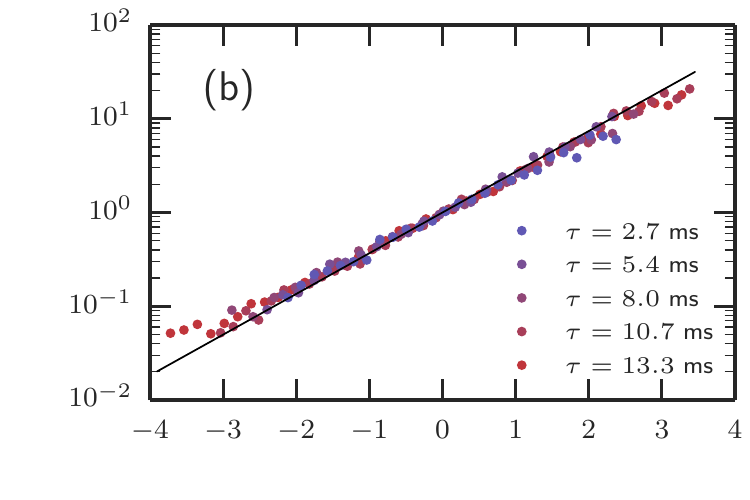}

	\includegraphics{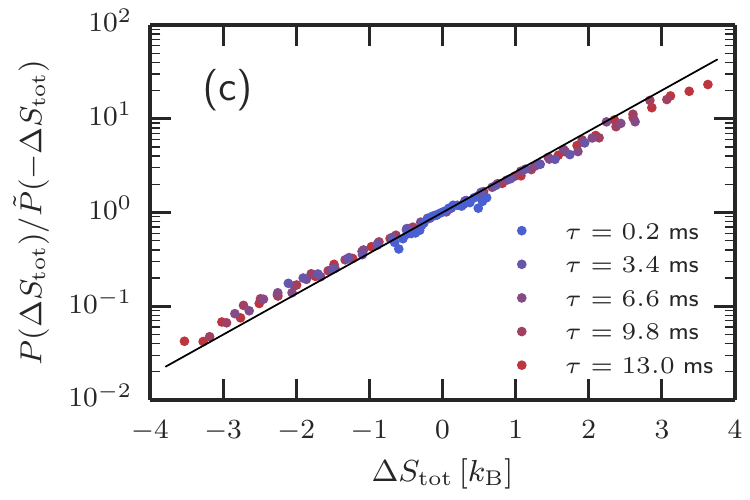}
	\includegraphics{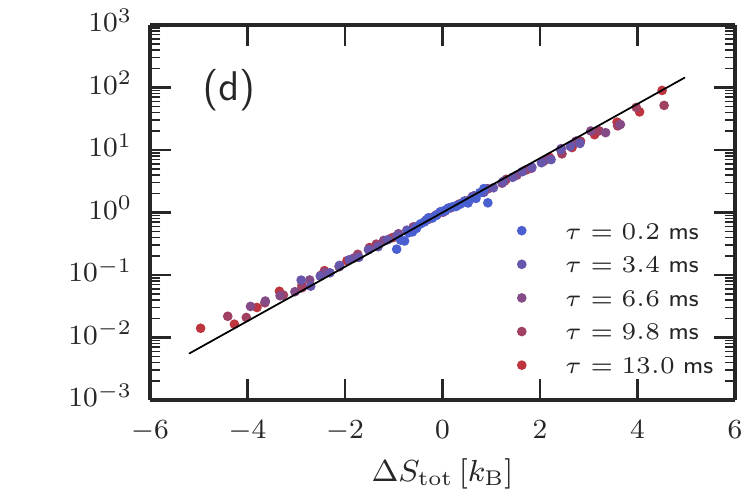}
	\caption{Ratio $\pf(\ep) / \pb(-\ep)$ on a logarithmic scale as a
		function of $\ep$ for different driving frequencies, amplitudes
		and durations $\tau$.  The black line corresponds to the theoretical
		prediction $\exp(\ep/\kb)$. Panel (a): Driving frequency
		$f_{\rm d} = 75\,\textrm{Hz}$, period $1/f_{\rm d} =
		13.3\,\textrm{ms}$, amplitude $q^*_{\rm max} =
		1.4\,\textrm{fC}$. Panel (b): Driving frequency $f_{\rm d} =
		75\,\textrm{Hz}$, period $1/f_{\rm d} = 13.3\,\textrm{ms}$,
		amplitude $q^*_{\rm max} = 2.3\,\textrm{fC}$. Panel (c):
		Driving frequency $f_{\rm d} = 1\,\textrm{kHz}$, period
		$1/f_{\rm d} = 1\,\textrm{ms} \approx RC$, amplitude $q^*_{\rm
		max} = 0.86\,\textrm{fC}$. Panel (d): Driving frequency $f_{\rm
		d} = 1\,\textrm{kHz}$, period $1/f_{\rm d} = 1\,\textrm{ms}
		\approx RC$, amplitude $q^*_{\rm max} = 1.3\,\textrm{fC}$. For
		panels (c) and (d), we considered durations up to 13 driving
		periods, hence $\tau_{\rm d} = 13\,\textrm{ms}$.}
	\label{f.res}
\end{figure}
Figure \ref{f.res} shows that the theorem (\ref{e.ft}) is verified with high
accuracy in our experiment.  This figure shows the ratio $\pf(\ep) / \pb(-\ep)$
between the distribution $\pf(\ep)$ of the entropy produced in the forward
process and of the distribution $\pb(-\ep)$ of (minus) the conjugated entropy
production $-\epb[\bar\traj]$ for the backward process for different driving
frequencies and amplitudes and for the durations.  Panels (a) and (b)
correspond to a driving frequency of $f_{\rm d} = 75\,\textrm{Hz}$ and hence a
driving period of $1/f_{\rm d} = 13.3\,\textrm{ms}$. The signal $q_t$ was
sampled at $20\,\textrm{kHz}$.  The driving amplitudes are $q^*_{\rm max} =
1.4\,\textrm{fC}$ for panel (a) and $q^*_{\rm max} = 2.3\,\textrm{fC}$ for
panel (b). The ensembles consist of $38\,424$ (panel a) and $38\,164$ (panel b)
trajectories.  Panels (c) and (d) correspond to a driving frequency of $f_{\rm
d} = 1\,\textrm{kHz}$ and hence the driving period is $1/f_{\rm d} =
1\,\textrm{ms} \approx RC$, which is the time constant of the circuit. Here, we
considered process durations up to 13 periods, and hence $\tau_{\rm d} =
13 / f_{\rm d} = 13\,\textrm{ms}$.
The signal $q_t$ was sampled at $100\,\textrm{kHz}$ and the
ensembles consist of $198\,144$ trajectories.  The driving amplitudes are
$q_{\rm max}^* = 0.86\,\textrm{fC}$ for panel (c) and $q_{\rm max}^* =
1.3\,\textrm{fC}$ for panel (d).  The black solid line corresponds to the
theory, $\exp\left( \ep / \kb \right)$. The fluctuation theorem (\ref{e.ft}) is
fulfilled with high accuracy (along up to four decades) for all the durations
considered.

\section{Conclusion}\label{s.ccl}

To summarize, in this work we have studied experimentally the
thermodynamics of the transition between two non-equilibrium states in
a parallel $RC$ circuit, in the light of the fluctuation theorem for
the entropy production (\ref{e.ft}). In such a situation, the
fluctuation theorem (\ref{e.ft}) presents a subtlety: for the backward
process it involves the distribution of the {\em conjugated entropy
production} (\ref{e.epbw}) rather than the distribution of the entropy
production.

We have characterized the statistics of the heat dissipation, entropy variation
and entropy production in the forward and backward processes. In particular,
for the backward process, we have studied the difference between entropy
production and conjugated entropy production. Furthermore, we have verified
that the detailed fluctuation theorem is fulfilled with high accuracy for
different driving frequencies and amplitude and for different process
durations, which confirms the universality of the result.

\section*{Acknowledgments}

L.G.~acknowledges the Max-Planck-Institut für Physik komplexer Systeme for its
hospitality.
L.G.~and É.R.~acknowledge financial support from Grant ENFASIS (FIS2011-22644,
Spanish Government).

\bibliographystyle{iopart-num}
\bibliography{bib}

\end{document}